\documentclass[showpacs,aps,twocolumn,pra,amsfonts,floatfix]{revtex4}
\usepackage{graphicx}
\usepackage{times}
\usepackage{color}
\usepackage[T1]{fontenc}
\usepackage{amssymb}
\usepackage{amsmath}

\usepackage[portuges]{babel}
\usepackage[latin1]{inputenc}

\newcommand{\be}{\begin{equation}}
\newcommand{\ee}{\end{equation}}
\newcommand{\br}{\begin{eqnarray}}
\newcommand{\er}{\end{eqnarray}}

\newcommand{\bd}{\begin{displaymath}}
\newcommand{\ed}{\end{displaymath}}

\newcommand{\bfig}{\begin{figure}}
\newcommand{\efig}{\end{figure}}

\def\3cdot{\cdot \cdot \cdot}

\def\om0{\omega _0}
\def\Om0{\Omega _0}

\def\text#1{{\rm{#1}}}

\def\->{\rightarrow}
\def\=>{\Rightarrow}
\def\-->{\longrightarrow}
\def\==>{\Longrightarrow}

\def\dag{\dagger}

\def\pr{^\prime}
\def\pr2{^{\prime\prime}}

\def\bfig{\begin{figure}}
\def\efig{\end{figure}}

\begin{document}

\title{A SIMPLE SCHEME FOR QUANTUM NON DEMOLITION OF PHONONS NUMBER OF THE NANOELECTOMECHANICS SYSTEMS}

\author{F. R. de S. Nunes$^{1}$, J. J. I. de Souza$^{1}$, D. A. Souza$^{1}$, R. C. Viana$^{2}$, e O. P. de S\'a Neto$^{1}$}

\affiliation{$^{1}$ Coordenação de Ciência da Computação,
Universidade Estadual do Piau\'i, CEP: 64202220, Parnaíba, Piau\'i, Brazil.}

\affiliation{$^{2}$ Centro Cir\'urgico do Hospital Dirceu Arcoverde, Parnaíba, Piau\'i, Brazil.}

\date{\today}

\begin{abstract}

In this work we describe a scheme to perform a continuous over time quantum non demolition (QND) measurement of the number of phonons of a nanoelectromechanical system (NEMS). Our scheme also allows us to describe the statistics of the number of phonons.

\end{abstract}

\maketitle

\section{QUANTUM MECHANICS MEASUREMENT PROBLEM}

In general, the measurement of an observable in a given quantum system disturbs its state, such that the observable variance is greater in a future measurement \cite{Milburn}. This is easily illustrated by a simple system, a harmonic oscillator of mass $m$ and momentum operator $p$ and in a thermal state, as previously considered by references \cite{meu0}-\cite{qnd}. It's possible to initially make a precise measurement in the $x$ position, the canonically conjugate operator to moment $p$. However, due to Heisenberg's uncertainty principle, $\delta p\geq\hbar/(2\delta x)$, and $p$ is disturbed. However, in an evolution following this measurement, $p$ induces a variation in $x$: $\dot{x}=[x,p^{2}/2m]/i\hbar$, resulting in, $x(t)=x(0)+p(0)t/m$. Therefore, using the uncertainty relation to calculate the uncertainty in $x$ for future measurements $(\delta x(t))^{2}\geq(\delta x(0))^{2}+(\hbar/2m\delta x(0))^{2}t^{2}$, we conclude that position and momentum are uncorrelated. The measurement apparatus acted randomly disrupting the observable being measured. 

\section{PROTOCOL TO MEASURE GENERAL QND}

The Quantum non demotion (QND) measurement is characterized as one that can be performed without disturbing the observable state. In a QND measurement, the observable $\mathcal{O}_{S}$ of the system $S$ is inferred by measuring an observable $\mathcal{O}_{A}$ of an auxiliary system $A$, without disturbing the next evolution of $\mathcal{O}_{S}$. After a finite number of successive steps the final state $S$ remains an eigenstate of $\mathcal{O}_{S}$.

Formally, if we have the total Hamiltonian:
\begin{eqnarray}
H&=&H_{S}+H_{A}+H_{I},
\end{eqnarray}
with $H_{S}$ being the system Hamiltonian, $H_{A}$ being the apparatus Hamiltonian, and $H_{I}$ being the Hamiltonian of the apparatus-system interaction. The QND measurement $O_{S}$ must satisfy the following properties: 
\begin{enumerate}
\item {$\frac{\partial H_{I}}{\partial O_{S}}\neq0$ and $[O_{A},H_{I}]\neq0$. This condition is because we want to measure $O_{S}$ through $O_{A}$. This implies the interaction Hamiltonian should be a function of $O_{S}$ and that $O_{A}$ varies accordingly, to interact with the system. In fact, this condition must be observed for any type of measurement, since it simply requires that the pointer system *CITAR* varies depending on the eigenvectors of the observable being measured;}
\item {The operator of the observable $O_{S}$ must commute with $H_{I}$. This observable can not be changed during the measurement process;}
\item {$\frac{\partial H_{S}}{\partial O_{S}^{C}}\neq0$. This is the main feature of QND measurement: after the interaction of $S$ with $A$ the conjugate observable $O_{S}^{C}$  is changed uncontrollably. So that this increase in variance does not affect the observable being measured, we have to demand that the Hamiltonian of the system does not depend on the conjugate observable. So a more restrictive way is to require $[H_{S},O_{S}]=0$, because then the observable being measured is a constant of movement.}
\end{enumerate}

\section{MODEL}

\begin{figure}[h]
\includegraphics[scale=0.5]{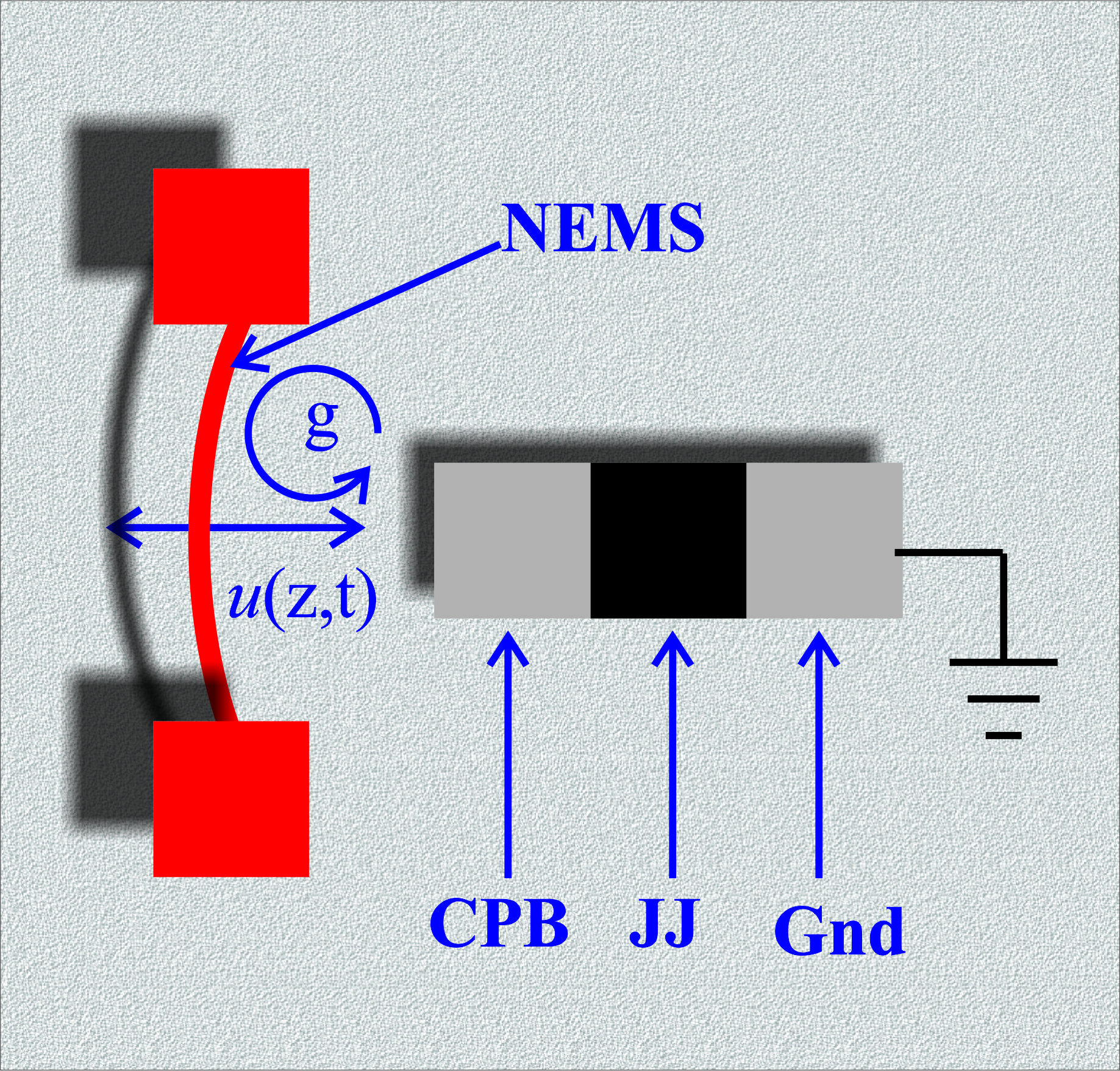}
\caption{Schematic Model.}
\label{Qubit+NEMS}
\end{figure}

The capacitive coupling between Quantum Bit (Qubit) and nanoelectromechanical system (NEMS) \cite{ML}-\cite{ML0} is illustrated in the figure \ref{Qubit+NEMS}. In quantum bit notation, the Hamiltonian of the Box of Cooper Pairs (\ref{55}) is written as 
\begin{equation}
H_{qb}=(E_{1}-E_{0})\sigma^{z}-\frac{E_{J}}{2}\sigma^{x},
\label{56}
\end{equation}
where $\sigma^{x}=\left|1\right\rangle\left\langle 0\right|+\left|0\right\rangle\left\langle 1\right|$, $\sigma^{z}=\left|0\right\rangle\left\langle 0\right|-\left|1\right\rangle\left\langle 1\right|$, and  $e$ is the eletron charge. $E_{n}=2E_{C}(n-n_{g})^{2}$ is the charging energy of $n$ cooper pairs, with $E_{C}=e^{2}/2C_{\sum}$, $C_{\sum}=C_{N}+C_{cpb}+C_{J}$. Also, $n_{g}=n_{N}+n_{cpb}$, where $n_{cpb}=C_{cpb}V_{cpb}/2e$ is the gate charge, $C_{cpb}$ is the capacitance and $V_{cpb}$ the potential difference of the Cooper pair box. $n_{N}=C_{N}V_{N}/2e$, is the gate charge, $C_{N}$ is the capacitance and $V_{N}$ is the potential difference of NEMS. $E_{J}$ is the capacitive energy of Qubit Josephson junction. Therefore, the necessary charging energy for the transition of one Cooper pair will be: 
\begin{eqnarray*}
E_{n+1}-E_{n}&=&2E_{C}\left[(n+1-n_{g})^{2}-(n-n_{g})^{2}\right],
\end{eqnarray*}
for $n=0$
\begin{eqnarray*}
E_{1}-E_{0}&=&2E_{C}(1-2n_{g})\nonumber\\
&&=2E_{C}(1-2n_{N}-2n_{cpb})
\end{eqnarray*}
Assuming small NEMS oscillation amplitude, we get the expression $C_{N}=C_{N}(0)+(\frac{\partial C_{N}}{\partial x})x$, with $x$ being the NEMS's flexion axis deformation position. Thus the capacitive interaction between the Qubit and NEMS mode is:
\begin{eqnarray}
H_{Q-N}&=&\hbar g \sigma^{z}(b+b^{\dag}),
\label{57} 
\end{eqnarray}
where,  $g=\sqrt{\frac{\hbar}{2m\omega}}\times[4n_{N}(0)E_{C}(\frac{\partial C_{N}}{\partial x})]/(\hbar C_{N})$, and  $\hbar$ is the Planck constant divided by $2\pi$.

The Complete Hamiltonian for this model is:
\begin{eqnarray} 
H_{\left|0\right\rangle,\left|1\right\rangle}&=&-\frac{E_{J}}{2}\sigma^{x}+\hbar\omega b^{\dag} b+\hbar g\sigma^{z}\left(b+b^{\dag}\right).
\label{58}
\end{eqnarray}

The Hamiltonian $H_{\left|0\right\rangle,\left|1\right\rangle}$ is written in the Cooper pair basis. However, changing the atomic basis to the new representation,
\begin{equation}
\sigma^{z}\rightarrow \sigma^{x}, \sigma^{x}\rightarrow -\sigma^{z},
\label{60}
\end{equation}
the Hamiltonian terms $H_{\left|0\right\rangle,\left|1\right\rangle}$ become:
\begin{equation}
H_{\left|-\right\rangle,\left|+\right\rangle}=\frac{E_{J}}{2} \sigma^{z}+\hbar\omega b^{\dag}b+\hbar g \sigma^{x}\left(b+b^{\dag}\right)
\label{61}
\end{equation}
with $\sigma^{x}=\left|+\right\rangle\left\langle -\right|+\left|-\right\rangle\left\langle +\right|$ and $\sigma^{z}=\left|-\right\rangle\left\langle -\right|-\left|+\right\rangle\left\langle +\right|$.

Making the rotation wave approximation to the Hamiltonian (\ref{61}), we have 
\begin{equation}
\tilde{\mathcal{H}}=\hbar\omega b^{\dag}b+\frac{E_{J}}{2}\sigma^{z}+\hbar g \left(\sigma_{-}b^{\dag}+\sigma_{+}b\right),
\end{equation}
where, $\sigma^{+}=\left|+\right\rangle\left\langle -\right|$, $\sigma_{-}=\sigma_{+}^{\dag}$, $\left|-\right\rangle$ is the fundamental atomic state, $\left|+\right\rangle$ is the excited atomic state.

For our case, considering a tightly dispersive regime, we can to expand the Hamiltonian with a Baker-Campbell-Hausdor as follows,
\begin{eqnarray*}
e^{-\lambda X}\tilde{\mathcal{H}}e^{\lambda X}&=&\tilde{\mathcal{H}}+\lambda\left[\tilde{\mathcal{H}},X\right]+\frac{\lambda^{2}}{2!}\left[\left[\tilde{\mathcal{H}},X\right],X\right]+\ldots
\end{eqnarray*}
where $\lambda=g/\Delta$, $\Delta=\omega-\nu_{a}$, $\nu_{a}=E_{J}/\hbar$ and $X=b^{\dag}\sigma_{-}+b\sigma_{+}$ results in the efective Hamiltonian 
\begin{eqnarray}
\mathcal{H}_{eff}&\approx&\hbar\left[\omega+\frac{g^{2}}{\Delta}\sigma_{z}\right]b^{\dag}b+\frac{\hbar}{2}\left[\nu_{a}+\frac{g^{2}}{\Delta}\right]\sigma_{z}.
\label{eff}
\end{eqnarray}
Now, with the equations of dynamics of the density operator,
\begin{eqnarray}
\dot{\rho}&=&\frac{-i}{\hbar}[H_{eff},\rho]+\kappa\mathcal{D}[b]+\gamma\mathcal{D}[\sigma_{-}]+\frac{\gamma_{\varphi}}{2}\mathcal{D}[\sigma_{z}]\nonumber\\
&=&\mathcal{L}\rho
\label{em}
\end{eqnarray}
where $\mathcal{D}[\alpha]=(2\alpha\rho\alpha^{\dag}-\alpha^{\dag}\alpha\rho-\rho\alpha^{\dag}\alpha)/2$.

\section{RESULTS}

\begin{figure}[t]
\includegraphics[scale=0.3]{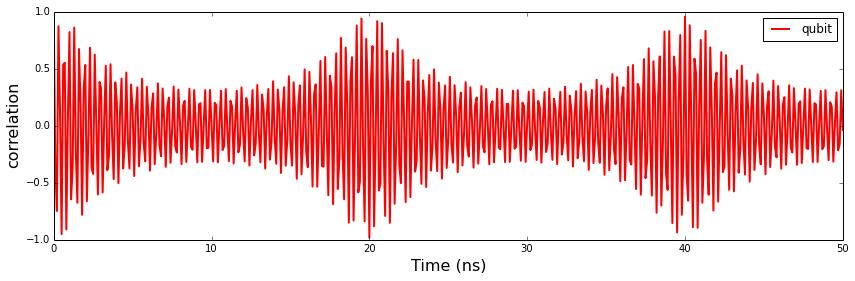}\\(a)\\
\includegraphics[scale=0.45]{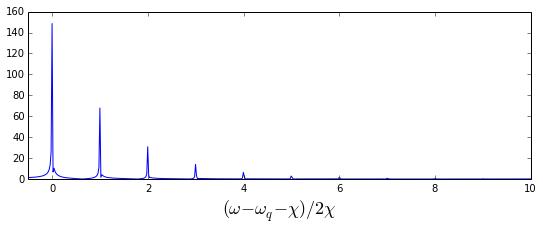}\\(b)\\
\includegraphics[scale=0.45]{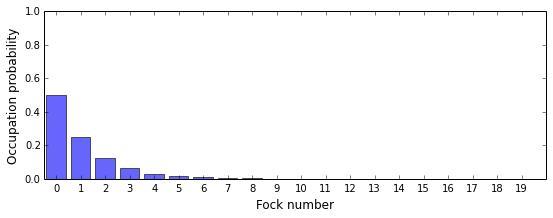}\\(c)\\
\caption{(a) Excited states correlation in time function, for $\chi=g^{2}/\Delta>>\kappa,\gamma$; (b) Qubit absorption spectrum given resolution number states of NEMS in termal state; (c) Visualization of the quantum states.}
\label{result0}
\end{figure}

With this we can calcule the correlation 
\begin{eqnarray}
\left\langle \sigma_{-}(t)\sigma_{+}(0)\right\rangle_{s}&=&Tr\left[\sigma_{-}e^{\mathcal{L}t}(\left|+\right\rangle\left\langle -\right|)\right],
\label{cor}
\end{eqnarray}
and finally the Qubit absorption spectrum.
\begin{eqnarray}
S(\omega)&=&\frac{1}{2\pi}\int_{-}^{}dte^{i\omega t}\left\langle \sigma_{-}(t)\sigma_{+}(0)\right\rangle_{s}.
\label{abspe}
\end{eqnarray}
We used the Qutip \cite{qutip} package to obtain numerical results for the correlation (fig. \ref{result0}.a), spectrum (fig. \ref{result0}.b), and its statistic distribution (fig. \ref{result0}.c). For our present calculation we used the Qubit in the excited state, and the NEMS in the vacuum state, with the number of thermal occupation of its reservoir being equal to one.

However, in this measurement protocol QND that measures the number of phonos, can to conduce the Qubit Stark frequence displacement in $\nu_{n}= \nu_{a}+ng^{2}/\Delta$, followed by the independent measure of Qubit state, once that the number of phonos is not changed in this process.

\section{DISCUSSION}

Motivated by a sete of discovery \cite{JG}-\cite{jg0}-\cite{ML}-\cite{ML0}-\cite{jg1}, we explored an electromechanical interaction in a highly dispersive regime in promoting for QND measurement scheme. We have demonstrated that the spectrum of the phonons of NEMS in the Qubit state resolution, thereby have access to each number of state and statistics of Bosen-Einsteis this ressoandor.

\section{Acknowledgements} 

Part of the calculations were performed with the Quantum Optics Toolbox. O. P. de S\'a Neto is grateful to Leonardo Dantas Machado for helpful discussions.


\end{document}